\documentclass[letterpaper,aps,pra,twocolumn,english,showpacs]{revtex4-1}

\PassOptionsToPackage{
	breaklinks=true,
	colorlinks=true,
	citecolor=	blue,
	linkcolor=	blue,
	urlcolor=	blue}{hyperref}

\usepackage{hyperref} 
\usepackage{amsmath}
\usepackage{amsfonts}
\usepackage{footnote}
\usepackage{todonotes}
\usepackage[caption=false]{subfig}
\usepackage{xcolor}
\usepackage{graphicx}
\usepackage[squaren]{SIunits}
\usepackage{IEEEtrantools}

\newcommand{\ket}[1]{\vert{#1}\rangle}

\begin{document}


\title{Testing Gravity with Cold-Atom Interferometers}

\author{G. W. Biedermann}

\altaffiliation[Present address: ] {Sandia National Laboratories, Albuquerque, NM 87123}

\author{ X. Wu}
\author{ L. Deslauriers}
\author{ S. Roy}
\author{ C. Mahadeswaraswamy}
\author{ M. A. Kasevich}
\email{kasevich@stanford.edu}
\affiliation{Physics Department, Stanford University, Stanford, CA 94305}
\date{\today}

\begin{abstract}
We present  a horizontal gravity gradiometer atom interferometer for precision gravitational tests.   The horizontal configuration is superior for maximizing the inertial signal in the atom interferometer from a nearby proof mass.  In our device, we have suppressed spurious noise associated with the horizonal configuration to achieve a differential acceleration sensitivity of 4.2$\times10^{-9}g/\sqrt{Hz}$ over a 70 cm baseline or 3.0$\times10^{-9}g/\sqrt{Hz}$ inferred per accelerometer. Using the performance of this instrument, we characterize the results of possible future gravitational tests.  We complete a proof-of-concept measurement of the gravitational constant with a precision of 3$\times10^{-4}$ that is competitive with the present limit of 1.2$\times10^{-4}$ using other techniques.  From this measurement, we provide a statistical constraint on a Yukawa-type fifth force at 8$\times$10$^{-3}$ near the poorly known length scale of 10 cm.  Limits approaching 10$^{-5}$ appear feasible.  We discuss improvements that can enable  uncertainties falling well below 10$^{-5}$ for both experiments. \end{abstract}

\pacs{06.30.Ft}

\maketitle

\section{Introduction}

Light-pulse atom interferometers demonstrate exceptional inertial sensitivity.  The nature of their construction lends long-term stability and intrinsic accuracy making them compelling candidates for advancing our knowledge of gravitational physics.  Recent work has shown the promise of this technology in a precision measurement of the gravitational constant \cite{Fixler2007, Rosi2014} as well as precision gradiometry \cite{McGuirk2002}, single-atom force sensors \cite{Parazzoli2012}, and navigation sensors \cite{Durfee2006,McGuinness2012, Butts2011, Geiger2011, Rakholia2014}.  New experiments aim to test the Weak Equivalence Principle by measuring the differential acceleration between atom species in a dual species accelerometer \cite{Dimopoulos2007, Schlippert2014} and future missions are being developed to deploy space-based gravity wave detectors \cite{Graham2013}. 

The weakness of the gravitational coupling presents a significant challenge for precision measurements of gravitational forces.  The 2010 CODATA values G with a relative standard uncertainty of $1.2\times10^{-4}$ \cite{Mohr2010}.  If taken as a steady trend, the uncertainty has improved by only one order of magnitude per century since the first measurements of G by Cavendish in 1798 \cite{Mills1979}.  High accuracy measurements following the first CODATA adjustment in 1986 disagreed with each other at the $10^{-3}$ level though their accuracies exceeded $10^{-4}$ \cite{Gillies1997}.  New understandings of systematic shifts in these measurements \cite{Kazuaki1999} and subsequent precision measurements have lead to the improved precision on G in 2010.  Therefore, an independent evaluation of G is welcome for determining the true value of G with greater accuracy.  In this paper, we present an atom interferometer that offers a new contribution to this endeavor with a forecast precision near $10^{-5}$.


In a related manner, inextricably linked are the precision of these measurements and the exploration of the dependence of gravity on the spatial separation of the participating test masses.  A myriad of theories predict departures from the Inverse Square Law (ISL) model just below the resolution of current experiments \cite{Adelberger2003}.  These theories often predict the existence of a new force mediated by a massive particle  exhibiting a characteristic range of $\lambda = \hbar/m_\gamma c$ where $m_\gamma$ is the particle mass.  In this case, the gravitational force would arise from a Yukawa potential of the form:
\begin{equation}\label{eq:Yuki}
U(r)= -\frac{\text{G} m_1 m_2}{r}\left(1+\alpha e^{-r/\lambda}\right),
\end{equation}
where $\alpha$ is the coupling strength, $m_1$ and $m_2$ are the two participating masses and $r$ is the spatial separation of the masses.  The work shown here offers the possibility to constrain $\alpha$ near $10^{-5}$ for $\lambda \sim$ 10 cm for an improvement of $10^2$ over current limits \cite{Hoskins1985, Moody1993, Yang2012}.

In this paper we present preliminary gravity tests using our technique with two experiments: a proof-of-concept measurement of the gravitational constant and a forecast of a statistical constraint upon a putative Yukawa-type fifth force.  Our promising results motivate further work to realize the full potential of this approach.  In the remainder of this paper we first discuss the atom interferometer measurement including the theoretical treatment along with our measurement technique  in \autoref{sec:theory}.  We then describe the apparatus and the current sensitivity in \autoref{sec:apparatus} followed by a characterization of the atom interferometer performance in \autoref{sec:sensitivity}.  Our results from a proof-of-concept measurement of G are found in \autoref{sec:G} and a forecast of a constraint upon the Yukawa potential is found in \autoref{sec:yukawa}. 

\section{Atom Interferometer Measurement}
\label{sec:theory}

The experiment  measures  acceleration using a pulsed-light, $\pi$/2-$\pi$-$\pi$/2 atom interferometer \cite{Berman1997}.  The functional principle of  the interferometer can be understood with a simple model.  This model encapsulates much of the behavior exhibited by the measurement process while neglecting several small yet important nuances such as the effect of magnetic fields, large local gradients in gravity and wavepacket overlap, as discussed below. 

Consider the case of a test mass undergoing constant acceleration.  A measurement of the position of the mass at three equi-spaced points in time defines the curvature or acceleration associated with its path as
\begin{equation}\label{eq:accel}
a=\frac{x_1-2x_2+x_3}{\mbox{T$^2$}}
\end{equation}
where T is the time between successive position measurements $x_i$.  For the atom interferometer, the test mass is the cesium atom and the position measurements are referenced to a pulsed, resonant optical field where the optical phase fronts act as the ticks of a ruler.  If this optical field is referenced to a stable frame, then the final interferometer phase shift reveals the acceleration of the atom with respect to that frame along the direction defined by the light propagation. Contributions from a constant velocity vanish in \autoref{eq:accel}.

The detailed theory of light pulse atom interferometry is available in Refs. \cite{Kasevich1992, Bongs2006}.  In brief, to perform these measurements, we interrogate the atoms with a velocity-sensitive, two-photon stimulated Raman transition coupling the 6S$_{1/2}$, F = 3 and F = 4 hyperfine ground states of atomic cesium.  These transitions imprint the optical phase associated with the Raman coupling onto the phase difference of the hyperfine ground state atomic wavefunctions.  This phase is a measure of the atom's position during the Raman pulse.  In the limit of short, resonant pulses, the transition rules between these two states take a simple form of 
\begin{eqnarray}\label{eq:tranRules}
\left|3,{\bf p}\right>  \rightarrow  e^{i\phi(t)} \left|4,{\bf p}+ \hbar {\bf k}_\text{eff}\right> \quad \nonumber \\
\left|4,{\bf p}+ \hbar {\bf k}_\text{eff}\right>  \rightarrow  e^{-i\phi(t)} \left|3,{\bf p}\right>,
\end{eqnarray}
where $\phi(t)={\bf k}_\text{eff}\cdot{\bf x}(t)$.  Here ${\bf x}$(t) is the mean position of the wavepacket at the pulse time t, ${\bf p}$ is the mean atom momentum and ${\bf k}_\text{eff}$ is the Raman wavevector defined by ${\bf k}_\text{eff}$ = ${\bf k}_{1}$ - ${\bf k}_{2}$, where ${\bf k}_{1}$ and ${\bf k}_{2}$ are the wavevectors of  two counter-propagating Raman beams.  Conservation of momentum dictates that the atomic momentum change by $\hbar {\bf k}_\text{eff}$ for an atom undergoing a Raman transition.  This amounts to a velocity change of $\approx$ 7 mm/s in our experiment, which leads to a macroscopic wavepacket separation of 0.6 mm over the duration of the interferometer which is  170 ms in this work.

\begin{figure}
\centering
\includegraphics[angle=0,width=8.5cm,keepaspectratio]{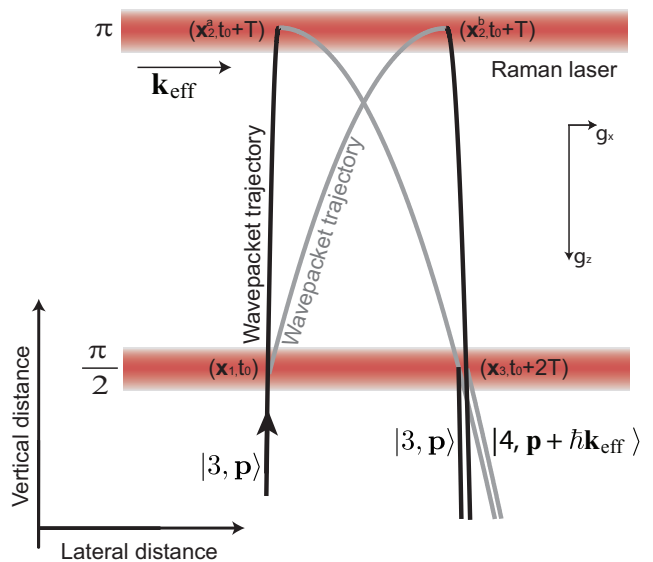}
\caption[Fountain recoil diagram]{Interferometer recoil diagram in an atomic fountain with a gravity gradient.  The prepared wave packet, $\ket{3,\bf{p}}$ interacts with the first $\pi$/2 pulse and splits in the lateral direction into a coherent superposition of states $\ket{3,\bf{p}}$ and $\ket{4,\bf{p}+\hbar\bf{k}_\text{eff}}$ while ascending to the apex of the trajectory.  At the apex the wave packets are redirected back toward one another by the $\pi$ pulse.  The final $\pi$/2 pulse recombines the wave packets near the original launch location.  The direction of $\bf{k}_\text{eff}$ determines that the interferometer measures $g_x$, the lateral acceleration. }
\label{fig:FountRecDia2}
\end{figure}

These Raman pulses drive coherent Rabi oscillations between the F = 3 and F = 4 ground states.  The pulses are characterized by the pulse area defined here as $\Theta \equiv \Omega_R \ \tau$, where $\Omega_R$ is the Rabi frequency which is assumed to be constant, and $\tau$ is the duration of the pulse.  As an example, for an atom initially in F = 3, a $\Theta = \pi$/2-pulse leaves the atom in an equal superposition of F = 3 and F = 4 analogous to an optical beam splitter.  It follows that a $\Theta = \pi$-pulse transfers an atom in F = 3 to F = 4 (and vice versa) corresponding to a mirror.  Therefore a $\pi$/2-$\pi$-$\pi$/2 pulse sequence creates a Mach-Zehnder style atom interferometer by splitting, redirecting and then recombining the atom wavepackets.  In practice, we employ an atomic fountain to loft atoms vertically upward and apply $\bf{k}_\text{eff}$ in the lateral direction (see \autoref{fig:FountRecDia2}).   The force of earth's gravitational pull causes the atoms to arc in a parabolic trajectory, returning them to the launch position such that they are in free-fall for the entire duration of the interferometer.  

\begin{figure*}[ht]
\centering
\includegraphics[angle=0,width=18cm,keepaspectratio]{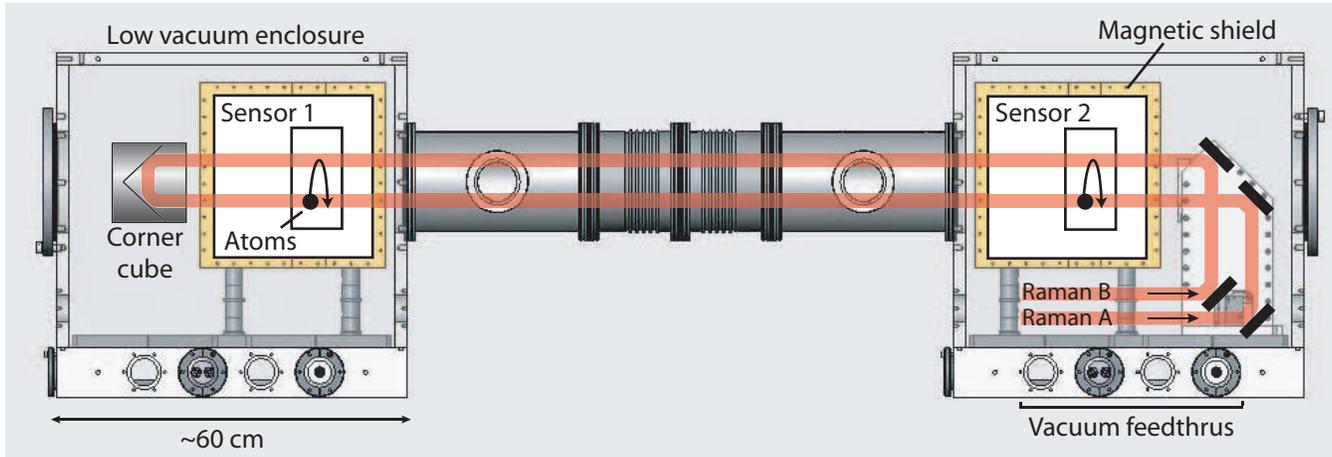}
\caption{A scaled schematic of the Raman laser delivery in the experiment (note that the sensor separation is reduced for the gravitational tests).  The Raman light enters the low vacuum enclosure through optical fiber vacuum feedthroughs.  Two collimated Raman beams counter-propagate in free space through both sensors and reflect from a corner cube giving two tiers for optical excitation.  The schematic shows the relative locations of the two UHV chambers in which the atom interferometers occur.  The fountain trajectories are exaggerated horizontally to depict motion.}
\label{fig:RD}
\end{figure*}

Each aforementioned position measurement is encapsulated in the phase $\phi(t)$.  Using the rules in Eq. \ref{eq:tranRules} for a $\pi$/2-$\pi$-$\pi$/2 interferometer, the transition amplitude for an atom beginning in state F = 3 is
\begin{equation}\label{eq:tranProb}
P\left(\left|4,{\bf p}+ \hbar {\bf k}_\text{eff}\right>\right)=\frac{1}{2}\left(1-\mbox{cos}(\Delta\phi)\right)
\end{equation}
where,
\begin{equation}\label{eq:deltaPhi}
\Delta\phi=\phi_1-\phi_2^a-\phi_2^b+\phi_3
\end{equation}
in analogy to \autoref{eq:accel}.  Here $\phi_i^j$ indicates the phase acquired during the i$^{\mbox{th}}$ pulse for path j.  For an atom in a uniform gravitational field it follows that
\begin{equation}\label{eq:phaseShift}
\Delta\phi = -{\bf k}_\text{eff} \cdot \textbf{\emph{g}} \mbox{T}^2 + \Delta\phi_0
\end{equation}
where $\textbf{\emph{g}}$ is the local acceleration due to gravity, T is the time between interferometer pulses and $\Delta\phi_0$ is an offset phase which vanishes when the measurement is referenced to a stable frame.

Additional effects contribute to the overall interferometer phase shift.  These include the phase evolution of the wavepackets in the two interferometer arms according to the Feynman path integral approach \cite{Feynman1965}, as well as a phase shift arising from imperfect overlap of these wavepackets following the final $\pi$/2-pulse.  These contributions are small relative to the light phase shift yet are important for high-accuracy metrology and are detailed in Refs. \cite{Bongs2006, Fixler2007}.


\section{Apparatus}
\label{sec:apparatus}

Two simultaneous acceleration measurements at different spatial locations typically constitute a gravity gradiometer. Such a measurement approximates the spatial rate of change in the gravity field. Accordingly, our gradiometer employs two spatially-separated gravimeters based on atom interferometry.  Each gravimeter is configured to measure the lateral component of gravity and the gravimeters are as well spatially separated in the lateral direction (see \autoref{fig:RD}) .  A key feature of this technique is that both gravimeters are interrogated with a common measurement laser which ideally propagates undisturbed between the gravimeters.  Since both gravimeters reference this laser, common mode platform noise is highly suppressed in the differential acceleration measurement, as discussed below.

Each gravimeter is a compact package with supporting opto-mechanical hardware densely arranged around an independent high vacuum chamber of $<$ 10$^{-9}$ Torr \cite{Biedermann2007}.  This package is surrounded by two layers of magnetic shielding to isolate the measurement.    To eliminate spurious noise associated with beam steering as discussed below, the entire gradiometer is enclosed in a low vacuum chamber of $\approx$ 50 mTorr.  The structure of the low vacuum chamber is carefully designed using finite-element-analysis to avoid significant misalignment of the Raman beams due to the large forces experienced by the structure from evacuation.

Using atomic fountain techniques, we prepare a 2.3 $\mu$K \footnote{This temperature is partially defined by the size of the detection aperture and the size of the detection laser beam due to the mapping of atom velocity to atom position at the end of the fountain.  From alternative measurements and theory we calculate a cloud temperature of 7 $\mu$K following launch.}, 3 mm 1/e$^2$ radius cloud of $\approx$ 10$^8$ cesium atoms in the 6S$_{1/2}$ F = 3, m$_f$ = 0 hyperfine ground state moving upward at 1 m/s. The atoms are in darkness during the fountain except for three temporally separated pulses of resonant Raman light which interrogate the trajectory as previously described.  Following the interferometer the atoms return back to approximately the same location from which they were launched. At this point acceleration information is encoded in the probability distribution between the two ground states.  In order to determine these two populations, and thus the probability distribution, we project the superposition and spatially separate the atoms according to their state with radiation pressure.  We then measure the respective populations of the two states with a simultaneous fluorescence detection technique described in Ref. \cite{Biedermann2009}.  Counting the number of atoms in both states enables the computation of a normalized population ratio which immunizes the result against shot-to-shot atom number fluctuations.     An alternative use of this apparatus as an atomic clock is presented in \cite{Biedermann2013}.

\begin{figure}
\centering
\includegraphics[angle=0,width=8.5cm,keepaspectratio]{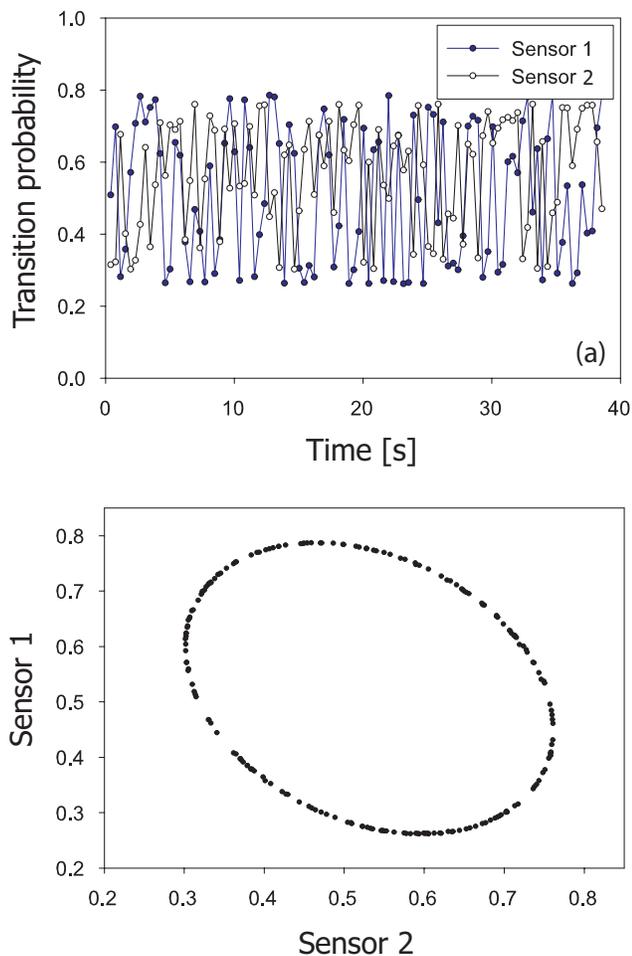}
\caption[Example of ellipse-based phase determination]{(a) An example of normalized transition probabilities from the two interferometers which comprise the gradiometer.  Common-mode noise in the optical delivery delivery system masks the phase information in the individual sensors while the difference phase is preserved. (b) Common-mode acceleration noise is suppressed when the accelerometer outputs are plotted parametrically.  200 data points form this example of a low phase noise ellipse.  The shot-to-shot fluctuations of the phase readout indicate a noise of 1.6 mrad per 20 point ellipse.}
\label{fig:GE}
\end{figure}

Due to the equivalence principle, it is impossible to distinguish between acceleration of the atoms and the reference mirror.  In practice, platform vibrations cause spurious phase shifts ($\Delta\phi_0 \neq 0$ in Eq. \ref{eq:phaseShift}) which severely limit the measurement sensitivity if not properly controlled \cite{Peters2001}.  In the present setup, this noise randomizes the phase of the interferometer at levels larger than $\pi$-radians.  However, our instrument uses two interferometers that share this noise in common such that the difference phase is preserved with high fidelity. Plotting the two transition amplitudes parametrically (see \autoref{fig:GE}) reveals a well-defined phase relationship between the sinusoidal outputs of the two interferometers characterized by the ellipticity \cite{Foster2002}.  Accordingly, we use ellipse-specific fitting to determine the differential phase and therefore the differential acceleration signal between the two sensors as discussed further in \autoref{sec:sensitivity}.

The Raman laser is sourced from a cavity-locked diode laser.  This system consists of a New Focus Vortex 6017 laser locked to an optical cavity via the Pound-Drever-Hall technique \cite{Fox2002}. The cavity is built from low-expansion Zerodur and has a hemispherical mirror geometry with a 10 cm separation and a finesse of 8000. The cavity length is piezo controlled and in this manner locked to a Cs resonance to eliminate drift and reduce low frequency acoustic noise.  The cavity output has a linewidth of $\approx$ 15 kHz and calculations show that the gradiometer noise floor associated with this laser is below the current sensitivity as is discussed below.

The scrubbed output from the cavity is fiber coupled and routed into the vacuum enclosures after further amplification and frequency control.  We use Photline fiber modulators to generate the required 9.1926 GHz hyperfine splitting frequency difference between the two counter-propagating Raman beams.  The final amplification is performed inside the low vacuum enclosure with an Eagleyard tapered amplifier.  The tapered amplifier output is spatially filtered then collimated to an r$_{1/e}$ = 6 mm beam waist and routed through a periscope and the two-level Raman beam configuration shown in \autoref{fig:RD}.  A corner cube reflector (PLX HM-25-1G) guarantees the parallelism of the two beam levels to within 5 $\mu$rad which is essential for good interferometer contrast.  In this configuration, the atoms interact with the first $\pi$/2-pulse immediately after the launch via the lower beam tier.  The second pulse ($\pi$-pulse) is applied with the upper beam tier at the apex of the fountain and the final $\pi$/2-pulse again uses the lower tier as the atoms travel down to the detection region.  A crossed linear polarization Raman excitation geometry is used to reduce susceptibility to parasitic reflections which give rise to standing wave AC Stark noise.

Using this apparatus, we observe continuous time records with a short term phase noise of 3.1 mrad/$\sqrt{Hz}$ inferred per interferometer.  For our system parameters, this corresponds to a differential acceleration sensitivity of 4.2 n$g$/$\sqrt{Hz}$ or 3.0 n$g$/$\sqrt{Hz}$ inferred per accelerometer. It is noteworthy that Bayesian techniques can be applied to the ellipse phase estimation routine to reduce the noise and systematic offset associated with simple ellipse fitting \cite{Stockton2007}.  

\begin{figure}
\centering
\includegraphics[angle=0,width=8.5cm,keepaspectratio]{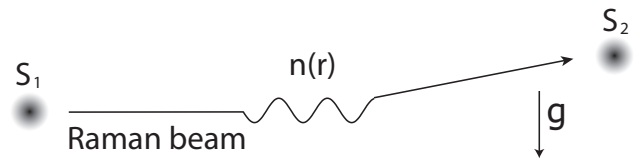}
\caption{Index of refraction variations in the air between the sensors (S$_1$ and S$_2$) result in an angular deviation of the Raman beam. Stochastic variations cause shot-to-shot fluctuations in the differential projection of the two measurement axes onto $\textbf{\emph{g}}$ which limits sensitivity.}
\label{fig:BD}
\end{figure}

\begin{figure}
\centering
\includegraphics[angle=0,width=8.5cm,keepaspectratio]{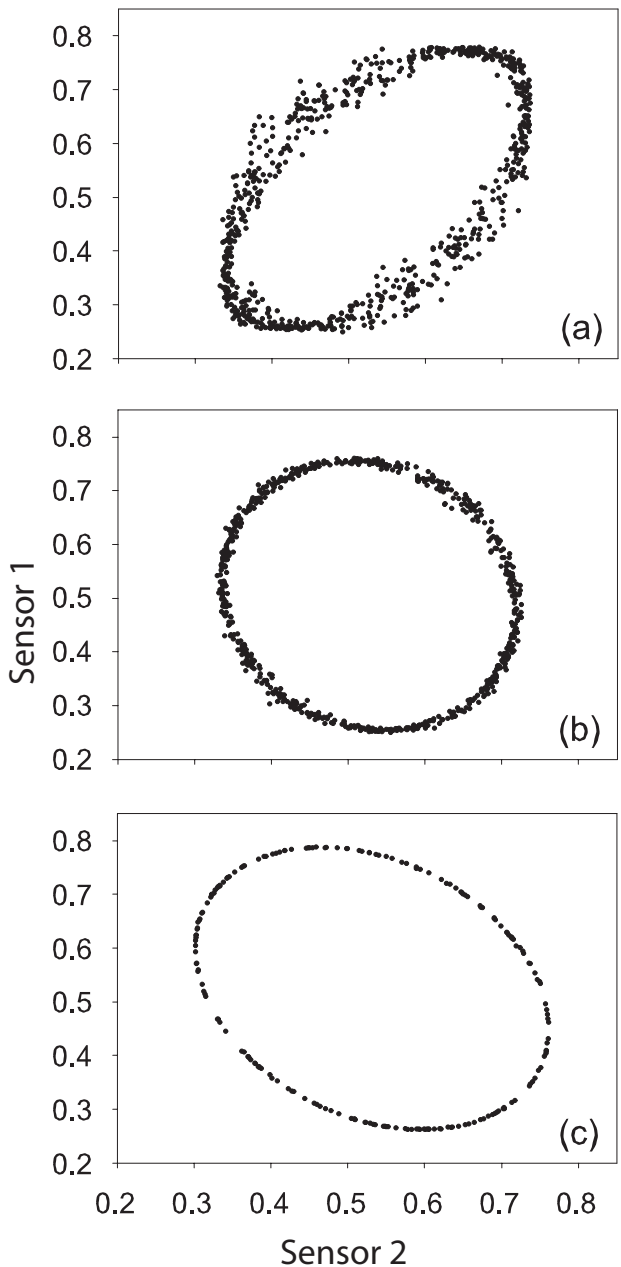}
\caption{Ellipse plots representing key noise limits. (a) Typical data with air between the sensors giving a differential phase noise of 190 mrad/$\sqrt{Hz}$. (b) Typical data after evacuating the air but generating the Raman beams with a DBR diode laser giving a differential phase noise of 38 mrad/$\sqrt{Hz}$. (c) Typical data using a cavity-locked laser giving a differential phase noise of 4.4 mrad/$\sqrt{Hz}$.}
\label{fig:EC}
\end{figure}

Although many parameters are explored to achieve this performance, two key experimental factors bear discussion here: intersensor beam steering effects and Raman laser frequency stability.  Perturbations to the Raman beam between the sensors produce a differential projection of each sensor's measurement axis onto $\textbf{\emph{g}}$ (see \autoref{fig:BD}). Considering that $|\textbf{k}_\text{eff}| \emph{g}_z$T$^2 \approx 10^6$ rad for typical instrument parameters, it is clear that beam steering at the nrad level will produce mrad interferometer phase shifts, commensurate with the device noise floor.  We find that in practice, stochastic index of refraction variations in the air between the sensors pose a severe limitation for horizontal gradient measurements such as this where ${\bf k}_\text{eff}$ is perpendicular to $\textbf{\emph{g}}$ (see \autoref{fig:EC}(a)).  In our system, this effect limits the differential phase noise to $>$ 190 mrad/$\sqrt{Hz}$.  Although phase readout below 1 mrad is routine in optical interferometers \cite{Bobroff1993}, heat sources in our apparatus such as magnetic field coils frustrate conventional solutions.  We find that enclosing the entire gradiometer in a low vacuum chamber eliminates the associated differential phase noise.

To a lesser degree, we find that Raman laser frequency noise limits the differential phase noise as shown in Ref. \cite{McGuirk2002} and later in Ref. \cite{Gouet2007}.  To illustrate this effect, consider that a discrete laser frequency change for one interferometer pulse results in a phase error of $\delta\phi = 4 \pi \delta\nu\ L/c$ where $\delta\nu$ is the laser frequency change, c is the speed of light and L is the separation distance of the two interferometers.  We have measured that for a mid-interferometer frequency jump of 1.161 MHz, a phase jump of 71.57 mrad results corresponding to an optical path length of L = 72.29 $\pm$ 0.09 cm after accounting for the effect of the vacuum windows.  This agrees with our physical measurement of 72.39 $\pm$ 0.25 cm.  In general, the interferometer phase noise is a function of the laser frequency noise spectrum up to a cutoff frequency commensurate with the Rabi frequency \cite{Gouet2007}.  We find that sourcing the Raman laser with a $\delta\nu\approx$ 1 MHz linewidth DBR diode limits the differential phase noise to 38 mrad/$\sqrt{Hz}$ as shown in \autoref{fig:EC}(b).  In contrast, a $\delta\nu\approx$ 15 kHz linewidth cavity-locked laser enables a noise of 3.1 mrad/$\sqrt{Hz}$ inferred per interferometer (see \autoref{fig:EC}(c)).  Calculations show that this cavity-laser performance is not a limit for the current device performance.

\section[Gradiometer sensitivity]{Gradiometer sensitivity}
\label{sec:sensitivity}

In this section, we present the current performance of the gradiometer including an evaluation of short and long term noise performance.  As previously discussed, the highly common-mode noise shared by the interferometers allows the use of ellipse-specific fitting to extract the differential phase signal between the two interferometers.  In our experiment, an optimal fit is typically achieved with 20 data points.  For a larger sample, the fit gains a susceptibility to slight drifts in detection offsets and interferometer phase during the collection of the ellipse points which typically takes 8 seconds for the 20 points.  We find that more than ten points are needed to achieve a good fit and at times corresponding to more than 100 points, system drifts  degrade the ellipse fit performance.

To determine the short term sensitivity of the interferometer we log a time record of the ellipse phase values with 20 points per ellipse and perform a double 3 sigma outlier cut on a dedrifted version of this record as briefly elaborated here.  To avoid erroneous results, we dedrift according to 20 ellipse phase averages.  We then calculate root-mean-square values of successive windows of 20 dedrifted phase points.  We remove 3 sigma outliers from this record according to the average rms value, then dedrift the data a second time and remove 3 sigma outliers again.  This protects the dedrift routine from the effects of very large outliers and the second cut typically removes much fewer points than the first.  

\begin{figure}
\centering
\includegraphics[angle=0,width=8.5cm,keepaspectratio]{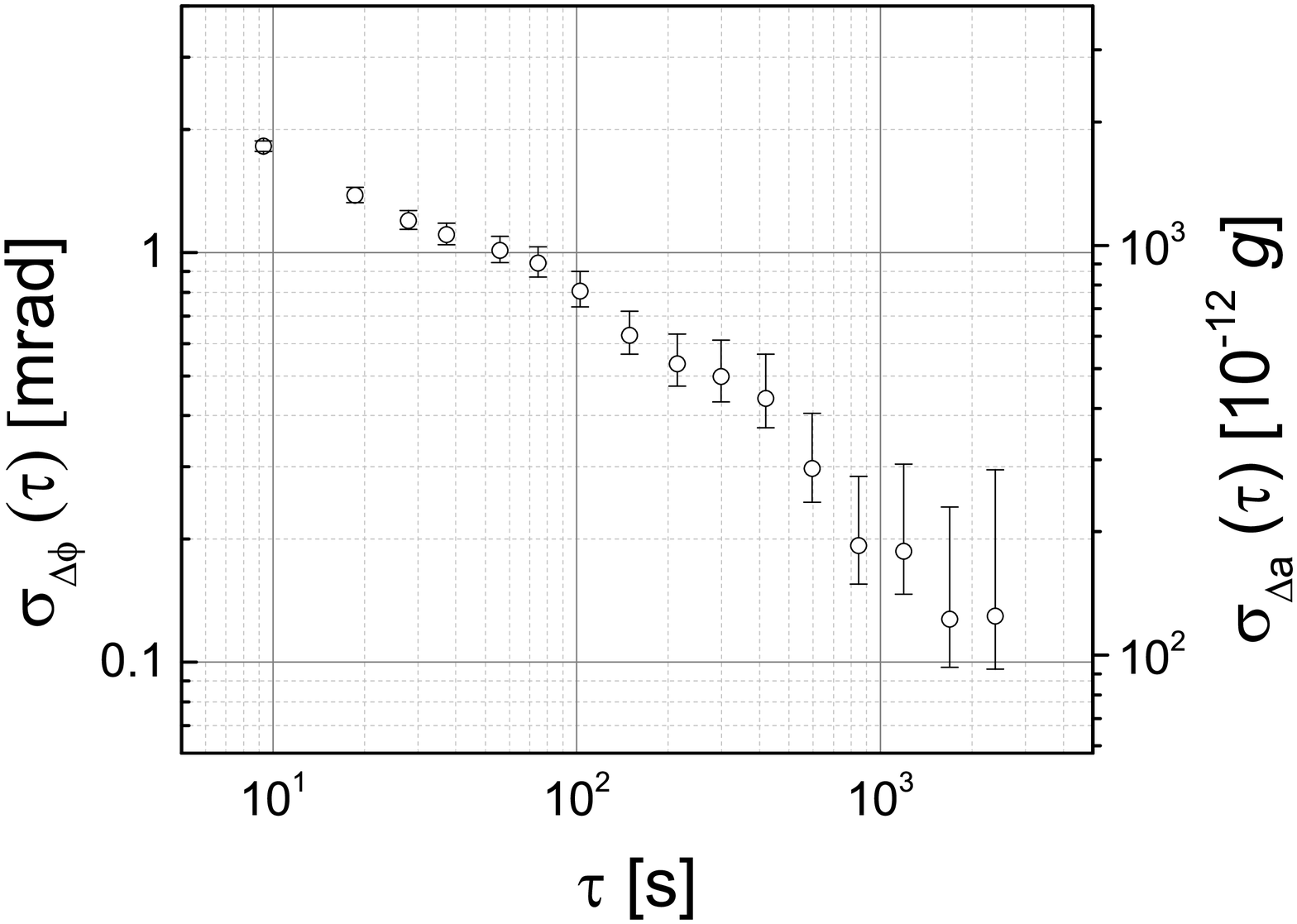}
\caption[Analysis of phase stability]{An Allan deviation analysis of the phase stability from the differential acceleration measurement shows that the system can integrate as white noise for periods of 2.5$\times 10^3$ seconds.  Here 0.1 mrad corresponds to 96 $\times10^{-12}$ $g$.  No attempt is made to correlate the data with system environmental parameters.}
\label{fig:BA}
\end{figure}

With this technique we  observe continuous time records with a short term noise of 1.6 mrad per ellipse.  In this case T = 85 ms and our repetition rate was 2.55 Hz giving a differential acceleration sensitivity of 4.2 $\times10^{-9} g /\sqrt{Hz}$ or 3.0 $\times10^{-9} g /\sqrt{Hz}$ inferred per accelerometer.  The long term performance shows white noise averaging for $2\times 10^3$ seconds (See \autoref{fig:BA}).    Electronic noise and noise caused by intensity and frequency fluctuations of the detection laser are negligible.  The long term noise is likely caused by environmental factors such as temperature drift.

\section{Gravitational Tests}
\label{sec:tests}

In this section we explore the suitability of the device for gravitational tests using a laboratory source mass.   We first show the instrument's viability for a precision measurement of G approaching 10$^{-4}$.  Second, we interpret this measurement as a test of the inverse-square law.  In both cases we provide an outlook for future gravity tests using atom interferometers.

\subsection[Gravitational Constant]{Gravitational Constant}
\label{sec:G}

\begin{figure}
\centering
\includegraphics[angle=0,width=8.5cm,keepaspectratio]{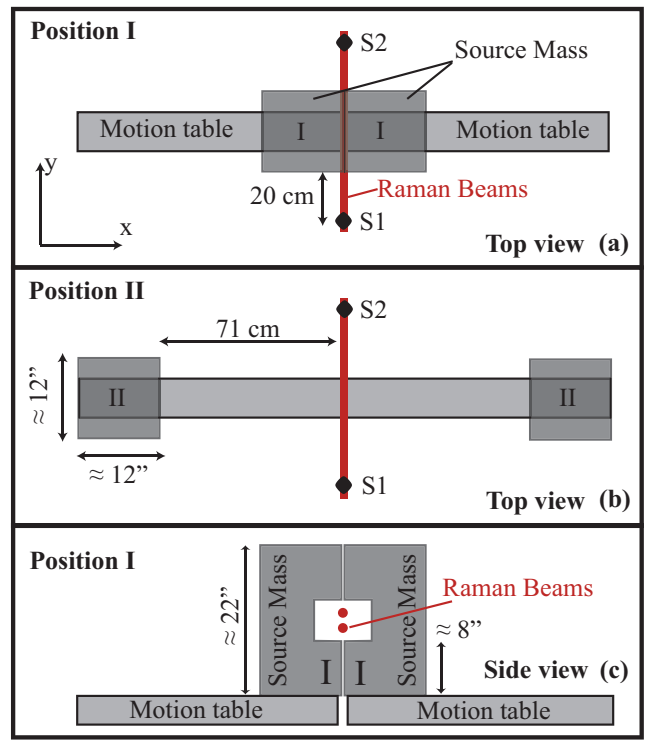}
\caption{Mass-sensor configuration for G measurement where S1 and S2 are the positions of the atom interferometer ensembles. The source masses are chopped between positions I and II, subfigures (a) and (b) respectively. A side view is shown in (c) depicting the $\approx$ 8" horizontal by $\approx$ 6" vertical opening to allow Raman beam propagation between the sensors.}
\label{fig:MS}
\end{figure}

To measure the gravitational constant, we take advantage of a symmetric source mass configuration to reduce sensitivity to atom-source positioning (see \autoref{fig:MS}). Relative positioning of the source mass and atoms is a significant source of error in previous measurements of G using atom interferometry \cite{Fixler2007}. By placing the source mass between the sensors, we make second-order the dependence of the field on many source position deviations.  For technical reasons, our experiment is performed with a small asymmetry in the distance of the two sensors from the source masses.  This does not inhibit the present demonstration as calculations show that our position repeatability of $<$ 5 $\mu$m is sufficient for a precision approaching 10$^{-5}$, nor do the results indicate the presence of any slow drifts in the mass signal.

\begin{figure}[h]
\centering
\includegraphics[angle=0,width=8.5cm,keepaspectratio]{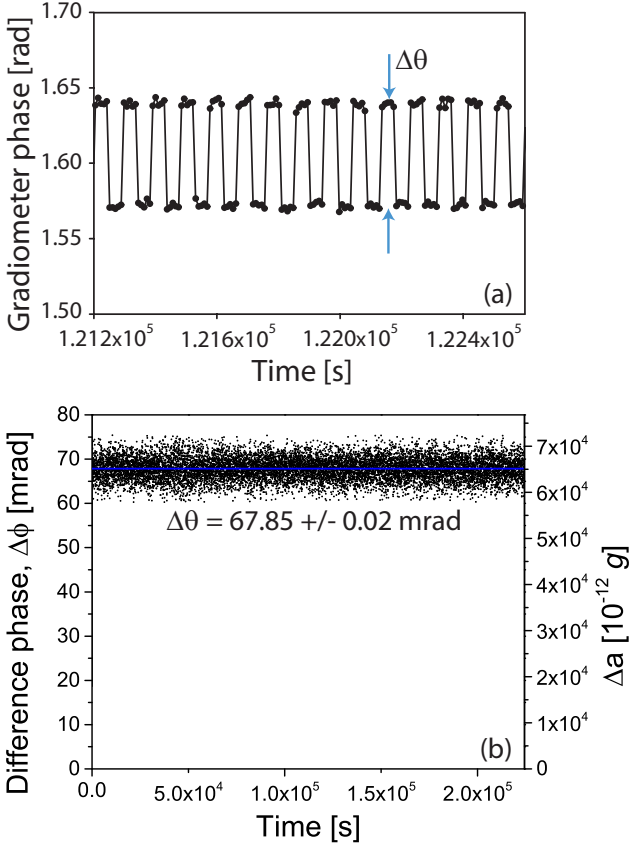}
\caption{(a) The gravity potential is chopped between two values to remove the sensitivity to long term drifts in phase.  This modulates the differential gravity vector along $\textbf{k}_\text{eff}$ by  (64.93 $\pm$ 0.02) $\times 10^{-9}$ $g$ at a repetition rate is 0.01 Hz. (b) Difference signal of the chopped gradiometer phase for a 2.6-day averaging interval.  The resulting phase shift is determined to be $\Delta\theta$ = 67.85 $\pm$ 0.02 mrad.}
\label{fig:Gsig}
\end{figure}

In the setup shown in \autoref{fig:MS} each of the two 540 kg source masses consists of 45 securely stacked laboratory lead bricks (2"x4"x8") strapped firmly to a LinTech 174630 precision positioning table. The positioning system enables rapid relocation of the source mass between the two end points or a 70 cm travel in less than 8 seconds.  The table, motors and drivers are specifically chosen to manage the torque and linear accelerations required for this motion profile. The positioning achieves this repeatability with simple mechanical limit switches at either end triggered by sloped flags. These switches are approached slowly at $\approx$ 1 mm/s to avoid overshoot due to the large inertia of the system. To modulate the field for the G measurement, the source masses are chopped between positions I and II at regular intervals (see \autoref{fig:MS}). The signal at each position is averaged for 40 s which is empirically chosen to minimize the introduction of noise from slow drifts in the gradiometer phase. The mass motion is synchronized with the interferometer timing system and data collection procedure. 

\begin{figure}[t]
\centering
\includegraphics[angle=0,width=8.5cm,keepaspectratio]{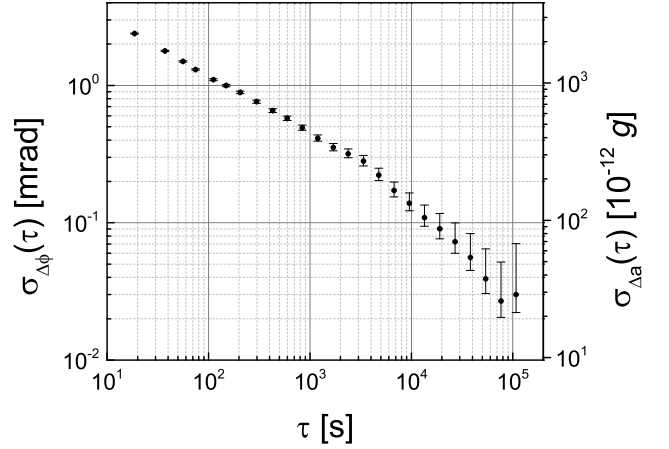}
\caption{Allan deviation of the difference phase. The local (3-point) de-drift algorithm results in a signature rise in the signal between 10$^2$ and 10$^4$ seconds. At longer times, the accuracy of the Allan deviation is restored giving an uncertainty of $\pm$ 0.02 mrad when extrapolated to the end of the data set, corresponding to 20 $\times$ 10$^{-12}$ $g$.}
\label{fig:adev}
\end{figure}

Using the technique described above, we measure the signal associated with modulating the gravity field between two values, giving a square wave output (see \autoref{fig:Gsig}(a)).  Slow systematic drifts contaminate this signal such that simple subtraction of adjacent values is inadequate to determine the wave amplitude.  We use three adjacent values to approximate the local linear rate of drift and largely remove this perturbation.  Explicitly, we report $\Delta\theta_k = \Phi^{II}_i - \frac{1}{2} (\Phi^{I}_{i-1} + \Phi^{I}_{i+1})$, where measurement $i$ is the average of 5 consecutive ellipse phase values and the superscript refers to the mass position in \autoref{fig:MS}.  Our simulations show that this analysis underestimates the short term noise by 13 \% but does not affect the interpretation of the long term sensitivity.  We remove occasional sections of data that are excessively noisy due to the loss of Raman laser cavity-lock. The resulting time records are concatenated as shown in \autoref{fig:Gsig}(b).  An Allan deviation of this record (see \autoref{fig:adev}) reveals that the brick chop signal integrates as $\tau^{-1/2}$ for 10$^5$ seconds.   Extrapolating the $\tau^{-1/2}$ trend to the full length of the data set gives a phase resolution on the gravitational square wave of $\Delta\theta$ = 67.85 $\pm$ 0.02 mrad.  This is equivalent to a resolution of 20 $\times$ 10$^{-12}$ $g$.  We therefore determine that this system can perform a measurement of the gravitational constant with a precision of $\delta$G/G = 3 $\times$ 10$^{-4}$.

This demonstrates that our system has the potential to produce a measurement of the gravitational constant competitive with the current precision of 1.2 $\times$ 10$^{-4}$ \cite{Mohr2010}. Achieving atom shot-noise limited sensitivity can enhance this result 20-fold \cite{Biedermann2009}.    Using a higher density material such as tungsten for the source mass, and arranging the source mass closer to the atoms with an optimized source mass geometry as discussed in \autoref{sec:yukawa} can increase the signal more than 6-fold. Furthermore, increasing the averaging time to 1 month can improve the result 3-fold.  Combining these possibilities, we forecast a precision of  $1 \times 10^{-6}$.  A unique possibility for the horizontal configuration is that the sensitivity can be further enhanced by increasing the interrogation time and extending the vertical dimension of the source mass, giving the potential for an additional order of magnitude improvement.  Finally, intrinsic sensitivity improvements via large momentum transfer atom optics \cite{McGuirk2000, Chiow2011} offer avenues for further investigation.

\subsection[Testing the Inverse Square Law]{Testing the Inverse Square Law}
\label{sec:yukawa}

This experiment may also be configured as a test of Newton's inverse square law (ISL) by directly measuring the spatial dependence of the gravitational field.  In this section we characterize a test that is possible with the current apparatus and then describe an optimized test using upgrades to the sensitivity and the mass configuration.

To place constraints on the strength and length scale of a Yukawa-type force, it is convenient to form ratio quantities in which both the absolute value of the mass as well as the gravitational constant cancel, leaving only the spatial dependence of the force law \cite{Fischbach1999}.  This avoids the necessity of comparison with the poorly known value of G and an absolute mass reference.  

In our experiment we measure relative quantities, chopping the test mass between a null reference position and a position of interest, to eliminate slow drifts in the interferometer phase.  We therefore construct the ratio 
\begin{equation}\label{eq:delta2}
\Delta \equiv \frac{(a_1-a_r)-(a_2-a_r)}{a_2-a_r} = \frac{a_1-a_2}{a_2-a_r},
\end{equation}
where $a_i$ are acceleration measurements performed at different positions and $a_r$ is a reference position.  In \autoref{eq:delta2}  the numerator and denominator quantities can be considered as two independent measurements with a statistical error equivalent to that described in \autoref{sec:G}.  Using this measured error, we predict the performance of a Yukawa test with our apparatus by forming  the constraint
\begin{equation}\label{eq:con}
\Delta_Y - \Delta_N \leq \sigma_m,
\end{equation}
where the subscripts $Y$ and $N$ refer to the Yukawa and Newtonian quantities respectively, and $\sigma_m$ is  calculated using error propagation of the measured 2$\sigma$ precision in \autoref{sec:G} or 40 $\times 10^{-12} g$.  We note that this precision was attained with a short, 2.6-day averaging duration which can in principle be increased.

We carefully choose the positions of the three measurements in order to optimize the constraint.  The reference measurement $a_r$  is taken at position II noted in \autoref{fig:MS}, while $a_1$ is taken at position I.  The optimal constraint on $\alpha$ occurs when $(a_2-a_r) = (a_1 - a_2)$.  This equates to locating the intermediate point $a_2$ at x = 21 cm which gives roughly half of the acceleration signal when compared to $a_1$.  Note that in this prediction the demonstrated experimental precision is reasonably assumed to hold at an intermediate point.

\begin{figure}
\centering
\includegraphics[angle=0,width=8.5cm,keepaspectratio]{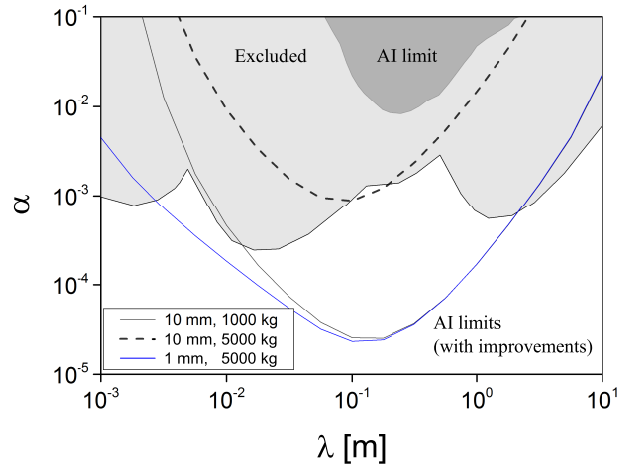}
\caption[Statistical Yukawa constraint]{Statistical atom interferometer (AI) Yukawa constraint using 2-$\sigma$ confidence bounds (shaded dark grey) is compared with the present limits from \cite{Hoskins1985, Moody1993, Yang2012} (shaded light grey).  This apparatus could be used to constrain $\alpha$ at the $8\times 10^{-3}$ level for $\lambda$ near 20 cm.  Three forecast curves are shown for the tungsten configurations detailed in \autoref{tab:params}.  With upgrades to the test mass configuration, the demonstrated sensitivity can exceed current limits with a source to atom distance of 10 mm and mass of 1000 kg.  Achieving atom shot noise limited sensitivity and increasing the mass to 5000 kg predicts limits approaching the $10^{-5}$ level for source to atom distances of 10 mm and 1 mm (see \autoref{fig:Yukd}). }
\label{fig:YC}
\end{figure}

For a Yukawa force, the acceleration is given by
\begin{equation}\label{eq:yuka}
a_i = \frac{\text{G} m}{r_i^2}\left( 1 + \alpha e^{-r_i/\lambda}\left(1+r_i/\lambda\right)\right).
\end{equation}
Using this, \autoref{eq:con} may be solved for $\alpha$ to determine an ISL constraint.  However, due to our complicated source mass geometry we numerically evaluate the terms in  \autoref{eq:con} for comparison with the value of $\sigma_m$ implied by our precision.  Specifically, in our experiment the Yukawa acceleration is given by:
\begin{equation}\label{eq:ay}
{a}_y^Y=\sum_i\frac{\text{G} m_i y_i }{r_i^{3}}\left(1+\alpha e^{-r_i/\lambda}\left(1+\frac{r_i}{\lambda}\right)\right),
\end{equation}
where $r_i = \left(x_i^2+y_i^2+z_i^2\right)^{1/2}$ while the Newtonian acceleration is given by 
\begin{equation}\label{eq:SMC}
a_y^N=\sum_i{\frac{\text{G} m_i y_i}{r_i^3}}
\end{equation}
where the subscript $i$ refers to a particular voxel in the mass distribution.  \autoref{fig:YC} shows parametric curves for which  \autoref{eq:con} would be satisfied for our device, along with the present limits from \cite{Hoskins1985, Moody1993, Yang2012}.  We predict a 2$\sigma$ statistical constraint on $\alpha$ of $8\times 10^{-3}$ with this apparatus.  This suggests that this experiment is currently within a factor of six of improving the limits on $\alpha$ near $\lambda=20$ cm.

\begin{figure}
\centering
\includegraphics[angle=0,width=8cm,keepaspectratio]{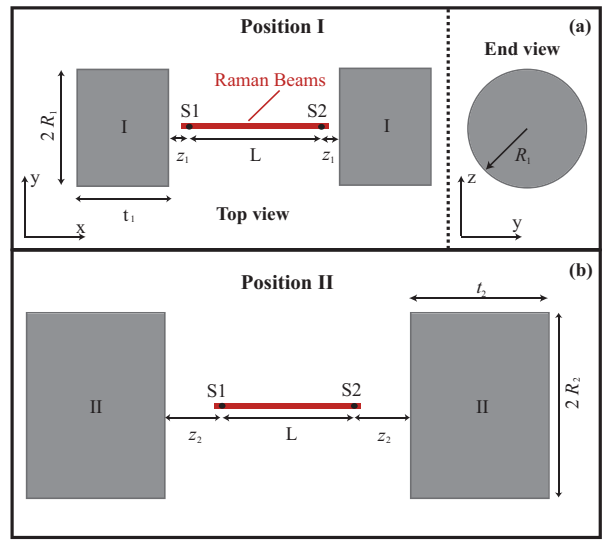}
\caption[Proposed mass-sensor configuration for an improved ISL test]{Proposed cylindrical mass configuration for an improved ISL measurement.  The source mass positions alternate between a null position and configuration I or II, (a) and (b) respectively.  The parameter $z_1$ represents the distance of closest approach to the atoms and L is chosen to be much larger than the spatial extent of the source masses.  Parameter values are given in \autoref{tab:params} for three configurations.}
\label{fig:Yukd}
\end{figure}

In future experiments, significant improvements to this constraint are possible.  We explore this by highlighting an optimized source mass geometry.  To begin, we note that the constraint is limited by the weakest of the two signals, $a_1$ and $a_2$, since the acceleration uncertainty is absolute.  In an experiment of this type, it is common practice to increase the source mass with increasing distance, to mute this effect \cite{Fischbach1999}. Furthermore, we consider an enhanced gravitational signal due to increased source mass density (tungsten instead of lead) and reduced proximity. The proposed setup is shown in \autoref{fig:Yukd}.  We choose a cylindrical source mass geometry to allow derivation of an analytic model.   For simplicity, this analysis assumes the ensemble is stationary in time.   

We use a bounded minimum search to find optimal values for all parameters shown in \autoref{fig:Yukd}.  These values are listed in  \autoref{tab:params} for three chosen cases: a mass limit of 1000 kg with a nearest approach of $z_1$ = 1 cm, and a mass limit of 5000 kg with a nearest approach of $z_1$ = 1 cm and 0.1 cm.   The results of these projections are shown in \autoref{fig:YC}. We note that the prediction for configuration A is readily achievable using an optimized geometry with the demonstrated sensitivity of the apparatus.   Plotted also are the constraints achievable  using configurations B and C, and our demonstrated atom shot noise limited detection \cite{Biedermann2009}.  Reducing the atom-mass proximity to $z_1$ = 0.1 cm significantly extends the constraint to shorter $\lambda$.   Further avenues for improvement as discussed in \autoref{sec:G} apply equally here prompting forecast exclusions of $\alpha$ below $10^{-5}$.

\begin{table}
\begin{tabular}{ | l | r | l | l | l | }
\hline
	\multicolumn{2}{ | c |}{Configuration}			& A		& B		& C		\\
  \hline
	Position I	& $z_1$ [cm] 	& 1.0		& 1.0 	& 0.1		\\
			& $t_1$ [cm]  	& 12.7	& 21.5 	& 21.3 	\\
			& $R_1$ [cm] 	& 13.6	& 22.5	& 21.4	\\
			& $m_1$ [kg]	& 117	& 660	& 588	\\
  \hline
	Position II	& $z_2$ [cm]	& 15.0	& 24.3	& 22.4	\\
			& $t_2$ [cm]	& 26.3	& 45.3	& 45.8	\\
			& $R_2$ [cm]	& 25.1	& 42.7	& 42.5	\\
			& $m_2$ [kg]	& 1000	& 5000	& 5000	\\
  \hline
\end{tabular}
\caption{Optimized tungsten source mass parameters for proposed ISL tests in \autoref{fig:Yukd} found by limiting the nearest source mass to atom ensemble distance to either 0.1 cm or 1.0 cm, and limiting the largest mass to either 1000 kg or 5000 kg.  Predictions using these parameters are shown in \autoref{fig:YC}.}
\label{tab:params}
\end{table}

Bringing the source to a distance of 0.1 cm from the atoms represents a significant experimental challenge as this is equivalent to the size of the cloud in the current apparatus.  However, recent progress in atomic fountains has demonstrated atom cooling and launch techniques that can be modified to achieve high localization and low expansion \cite{Dickerson2013}.  Recent theoretical work indicates that further refinements can provide a measurement at the 10 cm length scale exceeding well beyond the $10^{-5}$ level \cite{Wacker2010}.  At this proposed precision level, many sources of error can limit the accuracy.  Possibilities include edge effects from the finite source mass extent, surface flatness, and launch angle with respect to the source mass surface.  Furthermore, the extended baseline of L $\gg$ 1 m will place an additional constraint on the frequency stability of the Raman laser which scales with baseline.  Refinements  to both the source mass and source mass modeling will be necessary for these measurements.

\section{Conclusion}

We have presented here a horizontal gravity gradiometer for precision gravitational tests. Using this apparatus, we have demonstrated a proof-of-concept measurement of the gravitational constant with a precision of 3 $\times$ 10$^{-4}$, which is competitive with the present limit of 1.2$\times$10$^{-4}$. Improvements can enable uncertainties falling well below 10$^{-5}$. We have also interpreted this work as a constraint on a Yukawa-type fifth force and project a 10$^2$ improvement over current known constraints near $\lambda$ = 10 cm.  The horizontal configuration offers the potential for superior tests of gravitational physics.  The free-fall nature of the atom interferometer technique benefits from maximizing the inertially-relevant dwell time of the atoms near the proof mass. 
As a result, a surface oriented normal to gravity and probed in the same direction will achieve this goal.  However, this approach presents a new challenge in implementation, namely a first-order sensitivity to Raman laser beam steering which couples to the signal from earth's gravitational force.  We have shown that evacuation of the Raman beam path overcomes this challenge.  We also clearly show the importance of stabilization of the Raman laser frequency for low phase-noise measurements with meter-scale baselines.  Incorporating the former into a re-imagined test mass geometry as well as reducing the separation of the atoms and the proof masses can result in a significant improvement to our knowledge of gravity.

\begin{acknowledgements}

We are indebted to Kai Bongs, Matt Cashen, Jeff Fixler, Todd Gustavson, Ken Takase and Brent Young for countless contributions to the design and construction of the apparatus.  This work was supported by AFRL under Contract No. F19628-02-C-0096 and DARPA under Contract No. W911NF-06-1-0064.

\end{acknowledgements}

\bibliographystyle{apsrev4-1}
\bibliography{gravity}

\end{document}